\documentclass[twocolumn,prl,showpacs]{revtex4}

\usepackage{amsmath}
\usepackage{amsfonts}
\usepackage{amssymb}
\usepackage{epsfig}
\usepackage{dcolumn}
\usepackage{bm}

\newcommand{\s}{\sigma}

\newcommand{\al}{\alpha}

\newenvironment{definition}[1][Definition]{\begin{trivlist}\item[\hskip \labelsep
{\bfseries\small #1 1}]}{\end{trivlist}}

\begin{document}

\title{Fault-Tolerant Quantum Dynamical Decoupling}
\author{K. Khodjasteh$^{1}$ and D.A. Lidar$^{2,3}$}
\affiliation{$^{1}$Physics Department, University of Toronto, 60 St. George St., Toronto, Ontario, Canada
M5S 1A7}
\affiliation{$^{2}$Chemistry Department, University of Toronto, 80 St.
George St., Toronto, Otario, Canada M5S 3H6}
\affiliation{$^{3}$Current address: Chemistry and Electrical
  Engineering Departments, University of Southern California, Los
  Angeles, CA 90089}

\begin{abstract}
Dynamical decoupling pulse sequences have been used to extend
coherence times in quantum systems ever since the discovery of the
spin-echo effect. Here we introduce a method of recursively
concatenated dynamical decoupling pulses, designed to overcome both
decoherence and operational errors. This is important for coherent
control of quantum systems such as quantum computers.
For bounded-strength, non-Markovian environments, such as for the spin-bath that
arises in electron- and nuclear-spin based solid-state quantum computer proposals, we
show that it is strictly advantageous to use concatenated, as opposed to
standard periodic dynamical decoupling pulse sequences. Namely, the concatenated scheme is
both fault-tolerant and super-polynomially more efficient, at equal cost. We derive a
condition on the pulse noise level below which concatenated  is guaranteed
to reduce decoherence.
\end{abstract}

\pacs{03.67.-a, 02.70.-c, 03.65.Yz, 89.70.+c}

\maketitle

\textit{Introduction}.--- In spite of considerable recent progress, coherent control and quantum information
processing (QIP) is still plagued by the problems associated with
controllability of quantum systems under realistic conditions. The two
main obstacles in any experimental realization of QIP are (i)
\emph{faulty controls}, i.e., control parameters which are limited in
range and precision, and (ii) \emph{decoherence-errors} due to
inevitable system-bath interactions. Nuclear magnetic resonance (NMR)
has been a particularly fertile arena for the development of many
methods to overcome such problems, starting with the discovery of the
spin-echo effect, and followed by methods such as refocusing, and
composite pulse sequences \cite{Freeman:book}.  Closely related to the
spin-echo effect and refocusing is the method of dynamical decoupling
(DD) pulses introduced into QIP in order to overcome
decoherence-errors \cite{Viola:99,Zanardi:98b}. In standard DD one
uses a \emph{periodic} sequence of fast and strong symmetrizing pulses
to reduce the undesired parts of the system-bath interaction
Hamiltonian $H_{SB}$, causing decoherence. Since DD requires no
encoding overhead, no measurements, and no feedback, it is an
economical alternative to the method of quantum error correcting codes
(QECC) [e.g., \cite{Knill:99a,Steane:03,Terhal:04}, and references
therein], in the non-Markovian regime \cite{Facchi:04}.

Here we introduce \emph{concatenated} DD (CDD) pulse sequences, which
have a recursive temporal structure. We show both numerically and
analytically that CDD pulse sequences have two important advantages
over standard, periodic DD (PDD): (i) Significant fault-tolerance to
both random and systematic pulse-control errors (see
Ref.~\cite{Viola:02} for a related study), (ii) CDD is significantly
more efficient at decoupling than PDD, when compared at equal
switching times and pulse numbers. These advantages simplify the
requirements of DD (fast-paced strong pulses) in general, and bring it
closer to utility in QIP as a feedback-free error correction scheme.

{\it The noisy quantum control problem}.---
The problem of faulty controls and decoherence errors in the context
of QIP, as well as other quantum control scenarios \cite{Rabitz:00},
can be formulated as follows. The total Hamiltonian $H$ for the
control-target system ($S$) coupled to a bath ($B$) may be decomposed
as: $H=H_{S}\otimes {I}_{B}+{I}_{S}\otimes H_{B}+H_{SB}$, where $I$ is
the identity operator. The component $H_{SB}$ is responsible for
decoherence in $S$. We focus here on the single qubit case, but the
generalization to many qubits, with $H_{SB}$ containing only single
qubit couplings, is straightforward. We shall interchangeably use
$X,Y,Z$ to denote the corresponding Pauli matrices $\s_\al$, and
$\s_0$ to denote $I$. The system Hamiltonian is
$H_{S}=H_{S}^{\mathrm{int}}+H_{P}$, where $H_{S}^{\mathrm{int}}$ is
the intrinsic part (self Hamiltonian), and $H_{P}$ is an externally
applied, time-dependent control Hamiltonian. We denote all the
uncontrollable time-independent parts of the total Hamiltonian by
$H_{e}$, the \textquotedblleft else\textquotedblright\ Hamiltonian:
$H_e := H_{S}^{\mathrm{int}} + H_B + H_{SB}$. We
assume that all operators, except $I$, are traceless (traceful
operators can always be absorbed into $H_{S}\otimes {I}_{B}$,
${I}_{S}\otimes H_{B}$). We further assume that $\Vert H_{e}\Vert
<\infty $ \cite{norm}. Note that this is a physically reasonable
assumption, even in situations involving theoretically
infinite-dimensional environments (such as the modes of an
electromagnetic field), since in practice there is always an upper
energy cutoff \cite{foot1}. We consider \textquotedblleft
rectangular\textquotedblright\ pulses [piece-wise constant $H_{P}(t)$]
for simplicity; pulse shaping can further improve our results
\cite{Freeman:book}. An \emph{ideal} pulse is the unitary system-only
operator $P(\delta )=\mathcal{T}\exp [-i\int_{0}^{\delta}
H_{P}(t)dt]$, where $\mathcal{T}$ denotes time-ordering and $\hbar =1$
units are used throughout. A \emph{non-ideal} pulse, $U_{P}(\delta)=\mathcal{T}\exp [-i\int_{0}^{\delta}
\{H_{P}(t)+W_{P}(t)+H_{e}(t)\}dt]$, includes two sources of errors:
(i) Deviations $W_{P}$ from the intended $H_{P}$. Such deviations can
be random and/or systematic, generally operator-valued; (ii) The
presence of $H_{e}$ during the pulse.

{\it Periodic DD}.---
In standard DD one periodically applies a pulse sequence comprised of
\emph{ideal, zero-width} $\pi $-pulses representing a
\textquotedblleft symmetrizing group\textquotedblright\ $\mathcal{G}
=\{P_{i}\}_{i=0}^{|\mathcal{G}|-1}$ ($P_{0}=I$), and their inverses. Let
$\mathtt{f}_{\tau _{0}}=\mathcal{T}\exp [-i\int_{0}^{\tau
_{0}}H_{e}(t)dt]$ denote the inter-pulse interval, i.e., free
evolution period, of duration $\tau _{0}$.  The effective Hamiltonian
$H_{e}^{(1)}$ for the \textquotedblleft symmetrized
evolution\textquotedblright\ $\prod_{i=0}^{|\mathcal{G}|-1}P_{i}^{\dagger}
\mathtt{f}_{\tau _{0}}P_{i}=:e^{i|\mathcal{G}|\tau _{0}H_{e}^{(1)}}$ is given for
a single cycle by the \emph{first-order Magnus expansion}:
$H_{e}^{(1)}\approx
H_{\mathrm{eff}}=\frac{1}{|\mathcal{G}|}\sum_{i=0}^{|\mathcal{G}|-1}P_{i}^{\dagger} HP_{i}$
\cite{Viola:99}. This result is the basis of an elegant
group-theoretic approach to DD, which aims to eliminate a given
$H_{SB}$ by appropriately choosing $\mathcal{G}$
\cite{Viola:99,Zanardi:98b}. The \textquotedblleft universal
decoupling\textquotedblright\ pulse sequence, constructed from
$\mathcal{G}_{\mathrm{UD}}:=\{\s_0,\s_1,\s_2,\s_3\}$, proposed in
\cite{Viola:99}, plays a central role: it eliminates arbitrary
single-qubit errors. For this sequence we have, after using Pauli-group identities ($XY=Z$ and cyclic permutations), \texttt{p}$_{1}:=$ $e^{i\tau
_{1}H_{e}^{(1)}}=\prod_{i=0}^{3}P_{i}^{\dagger} \mathtt{ f}_{\tau
_{0}}P_{i}=\mathtt{f}_{\tau _{0}}X\mathtt{ f}_{\tau _{0}}Z \mathtt{f}_{\tau _{0}}X\mathtt{f}_{\tau _{0}}Z$, where $\tau _{1}=4\tau
_{0}$. The idea of dynamical symmetrization has been thoroughly
analyzed and applied (see, e.g., \cite{Viola:04,Facchi:04} and
references therein). However, higher-order Magnus terms can in fact
not be ignored, as they produce cumulative decoupling
errors. Moreover, standard PDD is unsuited for dealing with non-ideal
pulses \cite{Viola:02}.

{\it Concatenated DD}.---
Intuitively, one expects that a pulse sequence which corrects errors
at different levels of resolution can prevent the buildup of errors
that plagues PDD; this intuition is based on the analogy with
\emph{spatially}-concatenated QECC (e.g., \cite{Steane:03}). With this
in mind we introduce CDD, which due to its \emph{temporal} recursive
structure is designed to overcome the problems associated with PDD.

\begin{definition}
  \label{def1}
A concatenated universal decoupling pulse sequence:
$\mathtt { p}_{n+1}:=\mathtt{p}_{n}X\mathtt{ p}_{n}Z\mathtt{ p}_{n}X\mathtt{ p}_{n}Z$, where
$\mathtt{ p}_{0}\equiv \mathtt { f}_{\tau _{0}}$ and $n\geq 0$.
\end{definition}

Several comments are in order: (i) \texttt{p}$_{1}$ is the
\textquotedblleft universal decoupling\textquotedblright\ mentioned
above, but one may of course also concatenate other pulse sequences;
(ii) One can interpret \texttt{p}$ _{1}$ itself as a one-step
concatenation: \texttt{p}$_{1}:=$\texttt{p}$_{X}Y$\texttt{p}$_{X}Y$, where \texttt{p}$_{X}:=$ $\mathtt{f}$$X$$\mathtt{f}$$X$
($\mathtt{f}$$:=$$\mathtt{f}$$_{\tau _{0}}$) and Pauli-group identities
have been used.
(iii) Any pair, in any order, of unequal Pauli $\pi $-pulses can be
used instead of $X$ and $Z$, and furthermore a cyclic permutation in
the definition of \texttt{p}$_{1}$ is permissible; (iv) The duration
of each sequence is given by $T \lesssim \tau_n := 4^{n}\tau_0$ (after
applying Pauli-group identities); (v) The existence of a minimum pulse
interval $\tau _{0}$ and finite total experiment time $T$ are
practical constraints. This sets a physical upper limit on the number
of possible concatenation levels $n_{\max}$ in a given experiment
duration; (iv) Pulse sequences with a recursive structure have also
appeared in the NMR literature (e.g., \cite{Haeberlen:68Pines:86}),
though not for the purpose of reducing decoherence on {\em arbitrary}
input states.  We next present
numerical simulations which compare CDD with PDD.

{\it Numerical Results for Spin-Bath Models}.--- For comparing the performance of CDD vs PDD, we
have chosen an important example of solid-state decoherence: a spin-bath environment
\cite{Prokofev:00}. This applies, e.g., to spectral diffusion of an electron-spin qubit due to
exchange coupling with nuclear-spin impurities \cite{sousa:115322}, e.g., in semiconductor quantum
dots \cite{Burkard:99}, or donor atom nuclear spins in Si \cite{Kane:98}. Specifically, we have
performed numerically exact simulations for a model of a single qubit coupled to a linear
spin-chain via a Heisenberg Hamiltonian: $H_{e}=\omega _{S}\sigma _{1}^{z}+\omega
_{B}(\sum_{a=2}^{K}\sigma _{a}^{z})+\sum_{a>b\geq 1}^{K}j_{ab}\vec{\sigma} _{a}\cdot
\vec{\sigma}_{b}$. The system spin-qubit is labelled $1$; the second sum represents the Heisenberg
coupling of all spins to one another, with $j_{ab}=j\exp (-\lambda d_{ab})$, where $\lambda $ is a
constant and $ d_{ab}$ is the distance between spins.  Such exponentially decaying exchange
interactions are typical of spin-coupled quantum dots \cite{Burkard:99}. The initial state is a
random product state for the system qubit and the environment. The goal of DD in our setting is to
minimize (the log of) the \textquotedblleft lack of purity\textquotedblright\ of the system qubit,
$ l\equiv \log _{10}(1-\mathrm{Tr}[\rho _{S}^{2}])$, where $\rho _{S}$ is the system density matrix
obtained by tracing over the environment basis. At given CDD concatenation level $n$ we also
implement PDD by using the same minimum pulse interval $\tau _{0}$ as in CDD and the same total
number of pulses $N \lesssim 4^{n}$; this ensures a fair comparison. In all our simulations we have
set the total pulse sequence duration $T=1$, in units such that $(\omega _{S}T,\omega _{B}T,\lambda
d_{j,j+1})=(2,1,0.7)$.  Longer pulse sequences correspond to shorter pulse intervals $\tau _{0}$.
Note that we have chosen our system and bath spins to be similar species. Qualitatively, the number
of bath spins $K$ had no effect in the tested range $2\leq K\leq 7$, while quantitatively, and as
expected, decoherence rises with $K$. DD pulses were implemented by switching $H_{P}=h
\sigma_{1}^{\alpha}$, $\alpha \in \{x,z\}$, on and off for a finite duration $\delta >0$; note that $n\leq
n_{\mathrm{max}}(T,\delta )$.
We define the \emph{pulse jitter} $W_{P}$ as an additive noise contribution to $H_{P}$. It is
represented as $W_{P}^{\alpha}=\vec{r}^{\alpha}\cdot\vec{\sigma}_{1}$, with $\vec{r}^{\alpha}$
being a vector of random (uniformly distributed) coefficients. We distinguish between
\emph{systematic} ($W_{P}^{\alpha}$ fixed throughout the pulse sequence, but different for each
$\alpha $) and \emph{random} ($W_{P}^{\alpha}$ changing from pulse to pulse) errors.

\begin{figure}[tbp] \epsfig{figure=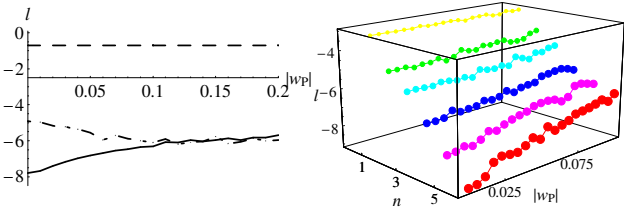,height=110mm}
  \vspace{-8.5cm}
\caption{Left: Performance of CDD (solid line, $n=4$) \textit{vs} PDD
(dot-dashed line) as a function of \emph{random} jitter fraction $|w_{P}|:=\left\Vert
W_{P}\right\Vert /\left\Vert H_{P}\right\Vert $, with pulse-width $\protect\delta
=10^{-5}T$, coupling strength $j=.2/T$ ($T$ is the total evolution time), and number of bath spins $K=2$, averaged over $90$ jitter
realizations. For comparison, the horizontal dashed line corresponds
to free evolution. Right: CDD as a function of random jitter
fraction $|w_{P}|$ and
concatenation level $n$. The vertical axis denotes $l=\log
_{10}(1-\text{purity})$ here and in Figs. 2 and 3.}
\label{fig1}
\end{figure}

\begin{figure}[tbp] \epsfig{figure=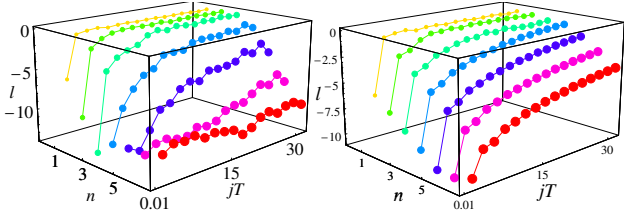,height=110mm}
  \vspace{-8.5cm}
\caption{{CDD (left) and PDD (right), as a function of system-bath
coupling $ j$; pulse width $\protect\delta =10^{-4}T$, number of bath
spins $K=5$, and without jitter ($W_{P}=0$). Note
the $l$-axis scale difference between CDD and PDD.}}
\label{fig2}
\end{figure}

\begin{figure}[tbp] \epsfig{figure=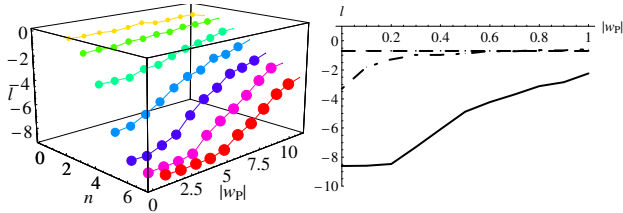,height=110mm}
  \vspace{-8.5cm}
\caption{{Left: CDD performance as a function of concatenation level
and \emph{systematic} jitter. The pulse width $\protect\delta
=10^{-4}T$, number of bath spins $K=5$,
$jT=15.0$, averaged over $7-80$ realizations (more realizations for
higher $n$). Right: CDD (solid line) vs
PDD (dot-dashed line) as a function of systematic jitter for $n=5$,
$\protect\delta =10^{-5}T$ , $K=5$, $j\protect\tau _{0}=3.0$, averaged
over $14-80$ realizations. The dashed line is pulse-free evolution. CDD
performance is unaffected up to $\sim 20\%$ jitter level.}}
\label{fig3}
\end{figure}

Our simulation results, shown in Figs.~\ref{fig1}-\ref{fig3}, compare
CDD and PDD as a function of coupling strength, relative jitter
magnitude, and number of pulses. Fig.~\ref{fig1}, left, compares CDD
and PDD at fixed number of pulses. CDD outperforms PDD in the random
jitter case with noise levels of up to almost $10\%$. Fig.~\ref{fig1},
right, shows the performance of CDD as a function of jitter magnitude
and concatenation level: the improvement is systematic as a function
of the number of pulses used. Figure~\ref{fig2} contrasts CDD and PDD
in the jitter-free case, as a function of system-bath coupling $j$. As
predicted in the analytical treatment below, CDD offers improvement
compared to PDD in decoherence reduction over a wide range of $j$
values.  Figure~\ref{fig3} compares CDD and PDD as a function of
systematic jitter. Superior performance of CDD is particularly
apparent. These results establish the advantage of CDD over PDD in a
model of significant practical interest, subject to a wide range of
experimentally relevant errors. We now proceed to an analytical
treatment.

\begin{figure}[tbp]
\epsfig{figure=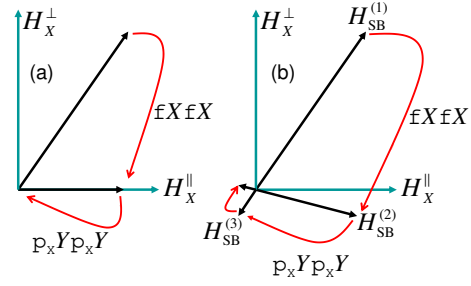,height=60mm}
\vspace{-2cm}
\caption{Projections involved in DD. (a) Perfect cancelation in
first-order Magnus case. (b) Extra rotation induced by higher-order
Magnus terms, and effect of concatenation.} \label{fig:proj}
\end{figure}

{\it Imperfect decoupling}.---
Consider DD pulse sequences composed of ideal pulses.  Let us
partition $H_{e}$ as $H_{e}=H_{X}^{\bot }+H_{X}^{\Vert} $, where $
H_{X}^{\bot} =Y\otimes B_{y}+Z\otimes B_{z}$ and $H_{X}^{\Vert}
=X\otimes B_{x}+H_{B}$. The super/subscripts $\bot ,X$ and $\Vert ,X$
correspond to terms that anti-commute and commute with $X\otimes
I_{B}$, respectively. Thus the effect of \texttt{p}$_{X}=$$\mathtt{f}X\mathtt{f}X$ in PDD can be viewed as a \emph{projection} of
$H_{SB}$ onto the component \textquotedblleft
parallel\textquotedblright\ to $X$, i.e., $ H_{X}^{\Vert}$. For the
$Y$ pulses in \texttt{p}$_{1}=$\texttt{p}$_{X}Y$\texttt{p}$_{X}Y$
we can similarly write $ H_{X}^{\Vert }=H_{Y}^{\bot }+H_{Y}^{\Vert }$,
where $\bot ,Y$ ($\Vert ,Y$) denotes anti-commutation (commutation)
with $Y$, whence $H_{Y}^{\Vert }=H_{B} $. Then the role of the $Y$
pulses is to project $H_{X}^{\Vert }$ onto $H_{Y}^{\Vert }$, which
eliminates $H_{SB}$ altogether, i.e., transforms $H_{e}=H_{SB}+H_{B}$
into a \textquotedblleft pure-bath\textquotedblright\ operator
$H_{B}$. This geometrical picture of two successive projections is
illustrated in Fig.~\ref{fig:proj}(a).  However, these projections are
imperfect in practice due to second-order Magnus
errors. Indeed, instead of a sequence such as $\mathtt{f}$$X$$\mathtt{f}$$X$, one has, after pulse $P_{i=X,Y}$, $I_{E,i}:=\exp [-i\tau
(H_{i}^{\Vert }-H_{i}^{\bot })]\exp [-i\tau (H_{i}^{\Vert}+H_{i}^{\bot })]$, where $H_{e}=H_{i}^{\Vert }+H_{i}^{\bot }$, and we
have accounted for the sign-flipping due to $P_{i}$.  Using the BCH
formula (e.g., \cite{Klarsfeld:89}), we approximate the total unitary
evolution as $ I_{E,i}=\exp [-i(2\tau )H_{\mathrm{eff},i}+O\left(
\lambda _{i}^{3}\right) ]$, where
$H_{\mathrm{eff},i}\!\!:=\!\!\mathcal{D}_{P_{i}}(\tau
)[H_{e}]=e^{-i\tau H_{i}^{\bot }/2}H_{i}^{\Vert }e^{i\tau H_{i}^{\bot}/2}$, where $\lambda _{i}^{3}:=\tau ^{3}\Vert H_{i}^{\bot }\Vert
^{2}\Vert H_{i}^{\Vert }\Vert $, and it is assumed that, since $\Vert
H_{e}\Vert <\infty $, one can pick $\tau $ such that $\lambda _{i}\ll
1$. The mapping $ \mathcal{D}_{P_{i}}(\tau )[H_{e}]$ clearly has a
geometric interpretation as a projection that eliminates $H_{i}^{\bot}$, followed by a rotation generated by $H_{i}^{\bot }$. \emph{This
rotation produces extra system-bath terms besides} $H_{i}^{\Vert
}$\emph{, hence imperfect DD}. This is illustrated in
Fig.~\ref{fig:proj}(b): in the first-order Magnus
approximation the transition from $H_{\mathrm{SB}}^{(2)}$ to
$H_{\mathrm{SB} }^{(3)}$ suffices to eliminate $H_{\mathrm{SB}}$,
i.e., $H_{\mathrm{SB} }^{(3)}=0$. But in the presence of
second-order Magnus errors $H_{ \mathrm{SB}}^{(3)}\neq
0$. The difference between CDD and PDD is precisely in the manner in
which this error is handled: in PDD the $H_{\mathrm{SB} }^{(3)}$ error
accumulates over time since the same procedure is simply repeated
periodically. However, in CDD the process of projection+rotation is
continued at every level of concatenation, as suggested in Fig.~\ref
{fig:proj}(b) (red arrow above $H_{\mathrm{SB}}^{(3)}$). In CDD,
$H_{\mathrm{ SB}}^{(m)}$ is shrunk with increasing $m$, in a manner we
next quantify.

{\it Convergence of CDD in the limit of zero-width pulses}.---
Decoupling induces a mapping on the components of $H_{e}$. For a
single qubit, writing $H_{e}=\sum_{\alpha=x,y,z}\sigma _{\alpha }\otimes
B_{\alpha }$, we have $H_{e}\overset{\mathtt{p}_{1}}{\mapsto }
H_{e}^{(1)}=\sum_{\alpha }\sigma _{\alpha }\otimes B_{\alpha }^{(1)}$,
where a second-order Magnus expansion yields:
$B_{0}^{(1)}=B_{0}$, $B_{x}^{(1)}=i\tau _{0}[B_{0},B_{x}]$, $
B_{y}^{(1)}=i\tau _{0}\frac{1}{2}([B_{0},B_{y}]-i\{B_{x},B_{z}\})$, $
B_{z}^{(1)}=0$. Let us define $\beta :=\left\Vert B_{0}\right\Vert$
and $J:=\max (\left\Vert B_{X}\right\Vert ,\left\Vert B_{Y}\right\Vert
,\left\Vert B_{Z}\right\Vert )$, where we assume $J<\beta <\infty $
\cite{foot2}.  Comparing with the model we have used numerically,
$J=O(\lambda)$ and $\beta=O(\omega_B)$. It is possible to show that a
concatenated pulse sequence $\mathtt{p}_{n}$ can still be consistently
described by a second-order Magnus expansion at all levels
of concatenation, provided the (sufficient) condition $\tau _{n}\beta
\ll 1$ is satisfied \cite{supp-mat}. We can then derive the recursive
mapping relations for $H_{e}^{(n-1)}\overset{\mathtt{p} _{n}}{\mapsto
}H_{e}^{(n)}=\sum_{\alpha }\sigma _{\alpha }\otimes B_{\alpha
}^{(n)}$:\ $B_{0}^{(n\geq 0)}=B_{0}$, $B_{x}^{(n\geq
\text{$1$})}=(i\tau _{n-1})[B_{0},B_{x}^{(n-1)}]$, $B_{y}^{(n\geq
\text{$2$})}=\frac{1}{2}(i\tau _{n-1})[B_{0},B_{y}^{(n-1)}]$,
$B_{z}^{(n\geq \text{$1$})}=0$. The propagator corresponding to the
whole sequence is $\exp (-i\tau _{n}H_{e}^{(n)})$, which in the limit
of ideal performance reduces to the identity operator. These results
for $B_{\alpha }^{(n)}$ allow us to study the convergence of CDD, and
bound the success of the DD\ procedure, as measured in terms of the
fidelity (state overlap between the ideal and the decoupled
evolution). This fidelity is given by \cite{Terhal:04}
\begin{equation}
f_{n}\approx 1-||\tau _{n}\widetilde{H}_{e}^{(n)}||^{2}\approx 1-(\tau _{n}h^{(n)}) ^{2}=:1-(\Phi _{\text{CDD}})^{2}, \label{eq:fid}
\end{equation}
where $\widetilde{H}$ is the system-traceless part of $H$, and $
h^{(n)}:=\max \{||B_{x}^{(n)}||,||B_{y}^{(n)}||\}$. We find that
\begin{equation}
\Phi_{\text{CDD}}\leq (\beta T/N^{1/2})^{n}(JT),
\label{eq:PhiCDD}
\end{equation}
where $T=N\tau _{0}\lesssim \tau _{n}=4^{n}\tau _{0}$ is the total
sequence duration, comprised of $N$ pulse intervals. In contrast,
$\Phi _{\text{PDD}}=Th^{(1)}$ yields
\begin{equation}
\Phi _{\text{PDD}}=2(\beta \tau _{0})(JT)=2(\beta T/N)(JT). \label{eq:PhiPDD}
\end{equation}
Note that for $N=4$, $\Phi_{\text{CDD}}=\Phi _{\text{PDD}}$ as
expected.  There is a physical upper limit to the number of
concatenation levels, imposed by the condition $\beta \tau _{n}\ll 1$.
Using this condition in the form $\beta =c/T$, where $c$ is some small
constant (such as $0.1$), and fixing the value of $ \beta $, we can
back out an upper concatenation level $n_{\max }=-\log _{4}
\frac{\beta \tau _{0}}{c}$; inserting this into Eq.~(\ref{eq:PhiCDD})
we have $\Phi _{\text{CDD}}\leq (c\beta \tau _{0})^{-\frac{1}{2}\log
_{4}\frac{ \beta \tau _{0}}{c}}(JT)$. We can now compare the CDD and
PDD bounds in term of the final fidelity:
\begin{equation}
\frac{1-f_{\text{CDD}}}{1-f_{\text{PDD}}}\leq \frac{(c\beta \tau
_{0})^{- \log _{4}\frac{\beta \tau _{0}}{c}}}{4(\beta \tau
_{0})^2}\overset{ \beta \tau _{0}\rightarrow 0}{\longrightarrow }0.
\label{eq:key}
\end{equation}
\emph{This key result shows that CDD converges super-polynomially
faster to zero in terms of the (physically relevant) parameter} $\beta
\tau _{0}$, \emph{at fixed pulse sequence duration}.  However, it is
important to emphasize that our bound on $\Phi _{\text{CDD}}$ is
unlikely to be very tight, since we have been very conservative in our
estimates (e.g., in applying norm inequalities and estimating
convergence domains).  Indeed, in our simulations (above) $\beta \tau_n
\approx 2$, which is beyond our conservatively obtained convergence
domain.

{\it Finite width pulses}.---
We now briefly consider the more realistic scenario of rectangular
pulses $\mathcal{T}\exp [-i\int_{0}^{\delta }\{H_{P}(t)+H_{e}(t)\}dt]$
of width $\delta \ll \tau _{0} $. In this case we can derive a
modified form of the condition $\beta \tau _{n}\ll 1$, required for
consistency (of using a second-order Magnus expansion at
all levels of concatenation) \cite{supp-mat}:
\begin{equation}
c\tau _{n}\beta +d\frac{\delta }{\tau _{n}}\ll 1, \label{eq:taunb-fw}
\end{equation}
where $c,d \sim 1$ are pulse sequence-specific numerical factors.  The
consistency requirement (\ref{eq:taunb-fw}) validates the analysis of
convergence of CDD for $\delta\neq0$, and we can reproduce the
advantage of CDD over PDD for $\delta =0$ [manifest in
Eq.~(\ref{eq:key})]. As expected Eq.~(\ref{eq:taunb-fw}) imposes a
more demanding condition on the total duration $\tau _{n}$, at fixed
bath strength $\beta $. While Eq.~(\ref{eq:taunb-fw}) cannot be called
a threshold condition (in analogy to the threshold in QEC), since it
depends on the total sequence duration, it does provide a useful
sufficient condition for convergence of a finite pulse-width CDD
sequence, and introduces the concept of error per gate which is
fundamental in QEC.

{\it Conclusions and outlook}.---
We have shown that concatenated DD pulses offer superior performance
to standard, periodic DD, over a range of experimentally relevant
parameters, such as system-bath coupling strength, and random as well
as systematic control errors. Here we have addressed the
\emph{preservation} of arbitrary quantum
states.  Quantum \emph{computation} can in
principle be performed, using CDD, over encoded qubits by choosing the
DD pulses as the generators of a stabilizer QECC, and the quantum
logic operations as the corresponding normalizer
\cite{ByrdLidar:01a,ByrdLidar:03}. Another intriguing possibility is
to combine CDD and high-order composite pulse methods \cite{Brown:04}.

\begin{acknowledgments}
Financial support from the DARPA-QuIST program (managed by AFOSR under
agreement No. F49620-01-1-0468) and the Sloan Foundation (to D.A.L.)
is gratefully acknowledged.
\end{acknowledgments}


\begin{thebibliography}{99}
\bibitem{foot1} This has been observed also in QECC work dealing with general error models \cite
{Knill:99a,Terhal:04}, but as has been pointed out there, if $\Vert H_{e}\Vert =\infty $ it must be
appropriately redefined.
\bibitem{foot2} This assumption is made to simplify some of our convergence arguments; while it
is not essential, it is reasonable, since we expect only a small number of bath degrees of freedom
to be coupled to a given qubit, whereas no restriction exists on the bath self-Hamiltonian
\end{thebibliography}

\end{document}